\documentclass[
    ,final            % use final for the camera ready runs
  ]
  {aipproc}

\layoutstyle{6x9}

\begin{document}

\title{Spontaneous CP violation and quark mass ambiguities}

\author{Michael Creutz}{
  address={Physics Department, Brookhaven National Laboratory, Upton,
NY 11973, USA}
}

\begin{abstract}
  I explore the regions of quark masses where CP will be spontaneously
  broken in the strong interactions.  The boundaries of these regions
  are controlled by the chiral anomaly, which manifests itself in
  ambiguities in the definition of non-degenerate quark masses.  In
  particular, the concept of a single massless quark is ill defined.
\end{abstract}

\maketitle

%%%%%%%%%%%%%%%%%%%%%%%%%%%%%%%%%%%%%%%%%%%%
%% MAINMATTER
%%%%%%%%%%%%%%%%%%%%%%%%%%%%%%%%%%%%%%%%%%%%

\section{Introduction}

In this talk I discuss two apparently distinct but deeply entwined
topics.  First, I ask for what quark masses is CP spontaneously
broken.  Second, I investigate whether a vanishing up quark mass is a
physically meaningful concept.  I should emphasize at the outset that
I will be exploring rather unphysical regions of parameter space.
This in some sense is a theorists fantasy, with no direct experimental
relevance.  Indeed, my real goal is to understand better how chiral
symmetry works in the strong interactions.  The main content of this
talk is contained in two recent papers
\cite{Creutz:2003xu,Creutz:2004fi}.  The basic ideas have roots in a
talk I gave in Como at the 1996 edition of this conference
\cite{Creutz:1996wg}.

I require a few assumptions.  First, the continuum limit of QCD should
exist and confine, with the only relevant parameters being the
coupling and the quark masses.  Then I will assume that chiral
symmetry is spontaneously broken in the usual way and that effective
chiral Lagrangians are qualitatively correct.  Finally I assume that
the anomaly removes any flavor singlet chiral symmetry.  In
particular, this implies that a single massless quark gives no exact
Goldstone boson.

The underlying concepts are all quite old.  In 1971 Dashen
\cite{Dashen:1970et} showed how CP symmetry could be spontaneously
broken in the strong interactions.  DiVecchia and Veneziano
\cite{DiVecchia:1980ve} observed the CP violation in chiral
Lagrangians at negative quark masses.  Georgi and McArthur
\cite{Georgi:1981be} showed that non-perturbative effects could give a
non-multiplicative shift to the up quark mass.  This motivated Kaplan
and Manohar \cite{Kaplan:1986ru} in their classic studies of
ambiguities in the up quark mass in the context of effective chiral
Lagrangians.  Banks, Nir and Seiberg \cite{Banks:1994yg} discussed the
fact that the concept of a vanishing up quark mass is not so clean.

\section{The effective meson theory}

I base my initial discussion on the effective theory for pseudoscalar
mesons in terms of a field taking values in the group $SU(3)$
\begin{equation}
\Sigma=\exp(i\pi_\alpha \lambda_\alpha/f_\pi)\ \in \ \hbox{ SU(3)}
\end{equation}
I work with the three flavor theory, i.e. I include the up, down, and
strange quarks.  The standard generators of the group SU(3) are given
by $\lambda_\alpha$, and the pseudoscalar octet fields are denoted as
$\pi_\alpha$.

Chiral symmetry is manifested in independent left or right global 
rotations on this field
\begin{equation}
\Sigma\rightarrow\ g_L^\dagger\ \Sigma\ g_R
\end{equation}
This symmetry is explicitly broken by the quark masses.  The lowest
order effective Lagrangian including the masses takes the form
\begin{equation}
L={f_\pi^2\over 4}{\rm Tr}(\partial_\mu \Sigma^\dagger \partial_\mu
 \Sigma) - v\ {\rm Re\ Tr}(\Sigma M)
\end{equation}
with the mass matrix
\begin{equation}
M=\pmatrix{ 
m_u & 0 & 0 \cr
0 & m_d & 0 \cr
0 & 0 & m_s \cr
}
\end{equation}
Expanding this density to quadratic order in meson fields and then
diagonalizing the resulting non-derivative term gives the usual result
that the meson masses squared are proportional to the quark masses,
including $m_{\pi^\pm}^2\ \sim \ m_u+m_d.$ For my purposes, I am
particularly interested in the isospin violation arising from the
up-down mass difference $m_d-m_u$.  This results in a mixing of the
{$\pi^0$} and {$\eta$} mesons, giving somewhat complicated formulae
for their masses
\begin{equation}
\matrix{
&m_{\pi^0}^2 \sim  
\ {2\over 3} \left(m_u+m_d+m_s
-\sqrt{m_u^2+m_d^2+m_s^2-m_um_d-m_um_s-m_dm_s}\right)\cr
&\cr
&m_{\eta}^2 \sim
\ {2\over 3} \left(m_u+m_d+m_s
+\sqrt{m_u^2+m_d^2+m_s^2-m_um_d-m_um_s-m_dm_s}\right)\cr
}
\end{equation}
Note in particular that it is possible to tune the parameters such
that $m_{\pi_0}^2$ vanishes.  This occurs when
\begin{equation}
m_u={-m_sm_d\over m_s+m_d}.
\label{boundary}
\end{equation}  
This vanishing mass does not require chiral symmetry.  It does,
however, occur at a somewhat unphysical location, requiring at least
one of the quark masses to be negative.

\section{Spontaneous CP violation}

Going to negative quark masses at first seems a bit strange, but in
such a regime unusual things do happen.  Note that because of the
anomaly, the signs of the quark masses can become significant.
Flavored chiral rotations can move the signs around, but the overall
sign of the determinant of the mass matrix is invariant.

\begin{figure}
  \includegraphics[height=.3\textheight]{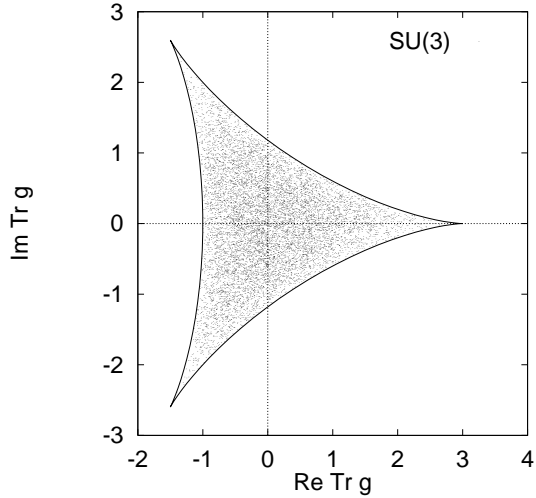}
\caption{The real and imaginary parts of the traces of 10,000 randomly
chosen $SU(3)$ matrices.  Note that the minimum real part occurs at
two distinct cube roots of unity.}
\label{su3}.
\end{figure}

The effective Lagrangian is useful for clarifying the expected
behavior with negative masses.  In the usual case with positive
masses, the vacuum involves quantum fluctuations about the maximum of
{${\rm ReTr}\Sigma$}.  This occurs at {$\Sigma=I$}.  However, now
consider the case of degenerate negative masses.  The vacuum instead
should occur at the minimum of ${{\rm ReTr}\Sigma}$.  The important
point is that $-I$ does not lie in the group $SU(3)$.  A simple
analysis shows that the minimum is doubly degenerate, occuring at
{$\Sigma=\exp(\pm 2\pi i /3)$}.  Fig.~\ref{su3} plots the traces of
10,000 random matrices to illustrate this result.  Note that CP
symmetry takes $\Sigma$ to $\Sigma^*$; thus, either of these solutions
involves a spontaneous breaking of CP.

\begin{figure}
  \includegraphics[height=.3\textheight]{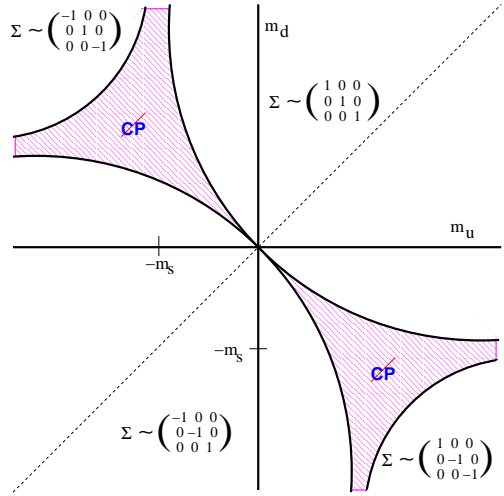}
\caption{The phase diagram of QCD as a function of the up and down quark
masses with a fixed positive strange quark mass.}
\label{cpr1}
\end{figure}

This CP violating phase can be approached continuously by passing
through the values of the quark masses in Eq.~\ref{boundary} where
$m_{\pi^0}^2$ vanishes.  Indeed, this equation represents the boundary
for the occurance of a pion condensed phase with $\langle \pi^0
\rangle\ne 0$.  Similar boundaries occur at the appropriate branches
of
\begin{equation}
m_u={-m_sm_d\over \pm m_s\pm m_d}
\end{equation}
The full phase diagram is sketched in Fig.~\ref{cpr1}.

In the CP violating phases the vacuum fluctuates about non-trivial
complex matrices of form
 \begin{equation}
\Sigma=\pmatrix{
e^{i\phi_1}&0&0\cr
0&e^{i\phi_2}&0\cr
0&0&e^{-i\phi_1-i\phi_2}\cr
}
\end{equation}
where the angles satisfy
\begin{equation}
m_u \sin(\phi_1)=m_d\sin(\phi_2)=-m_s\sin(\phi_1+\phi_2)
\end{equation}
The CP violating phases are separated from the conserving ones by
second order transition lines occuring when $m_{\pi^0}=0$.  The former
have two degenerate vacua related by $\phi_i\leftrightarrow -\phi_i$.

\section{Ambiguities in the up quark mass}

\begin{figure}
  \includegraphics[height=.3\textheight]{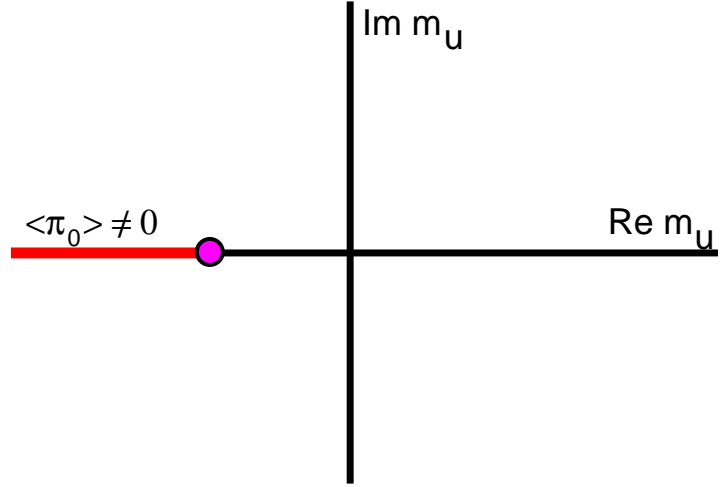}
\caption{The qualitative phase diagram as a function of a complex up
quark mass with fixed strange and down quark masses.  (Here I ignore
the reappearance of a CP conserving phase at large negative up quark
mass.)}
\label{oneflavor}
\end{figure}

I now take a section through this diagram.  Phenomenologically, the
down and strange quark masses appear to definitely not vanish.
Consider fixing them at some positive values and study the dependence
of the theory as a function of the up quark mass alone.  Thus follow a
horizontal line at fixed $m_d$ in Fig.~\ref{cpr1}.  To enable
continuation around the boundary of the CP violating phase, extend the
$m_u$ dependence into the complex plane.  This gives the qualitative
structure sketched in Fig.~\ref{oneflavor}.

The expectation is a first order transition along negative ${\rm Re}\
m$ axis.  This ends at a second order critical point at non-zero ${\rm
Re}\ m<0$.  Along the first order line there is a spontaneous
breaking of CP.  The transition has a simple order parameter $\langle
\pi_0 \rangle$.  The presence of the gap below $m_u=0$ and the CP
violating phase were noted some time ago by Di Vecchia and Veneziano
\cite{DiVecchia:1980ve}.

Note that nothing significant occurs at $m_u=0$ when $m_d\ne 0$.  This
raises an interesting question: Does $m_u=0$ have any physical
significance?  I now argue that this is not a well posed question if
$m_d\ne 0$ and $m_s\ne 0$.  One consequence of this observation is
that a vanishing up quark mass is an unacceptable solution to the
strong CP problem.

A crucial message here is that the concept of an ``underlying basic
Lagrangian'' does not exist.  Field theory is full of divergences that
must be regulated.  It is only the underlying symmetries of the theory
that remain significant.  The case of a single massless quark gives no
special symmetry because of the anomaly.  Unlike the multiple
degenerate quark case, no exact Goldstone bosons should appear at
$m_u=0$.

\subsection{The renormalization group trajectory}

A continuum field theory is defined as a limit of a cutoff theory.
For QCD the bare parameters are the coupling {$g$} and quark masses
{$m_i$}.  These must be renormalized for the continuum limit; indeed
both bare parameters renormalize to zero in a well defined way given
by the renormalization group equations.  For this discussion I denote
the cutoff as a minimum length $a$.  This corresponds to the inverse
of a large momentum scale.  To simplify the discussion I will consider
a single quark mass.  Then the well known flow equations in the small
coupling limit take the form
\begin{equation}\matrix{
&a{d\over da}g=\beta(g)=\beta_0 g^3+\beta_1 g^5 +\ldots
+\hbox{ non-perturbative}\cr
&a{d\over da}m=m\gamma(g)=m(\gamma_0 g^2+\gamma_1 g^4 +\ldots)
+\hbox{ non-perturbative}\cr
}
\end{equation}
The first few coefficients $\beta_0,\ \beta_1$, and $\gamma_0$ are
scheme independent.  In these equations the ``non-perturbative'' parts
fall faster than any power of $g$ as $g\rightarrow 0$.  Q crucial
point, to which I will return, is that these contributions are not
proportional to the quark mass.

The solution to these equations is standard
\begin{equation}\matrix{
&a={1\over \Lambda} e^{-1/2\beta_0 g^2} g^{-\beta_1/\beta_0^2}
(1+O(g^2))\cr
&m=Mg^{\gamma_0/\beta_0}
(1+O(g^2))\cr
}
\end{equation}
Rewriting shows how the coupling and mass go to zero in the
continuum limit $a\rightarrow 0$
\begin{equation} \matrix{
&g^2\sim {1\over \log(1/\Lambda a)}\rightarrow 0\cr
&m\sim M\ \left({1\over \log(1/\Lambda
a)}\right)^{\gamma_0/2\beta_0}\rightarrow 0
}
\end{equation}
The first part of this equation represents the famous phenomenon of
``asymptotic freedom.''

\begin{figure}
  \includegraphics[height=.3\textheight]{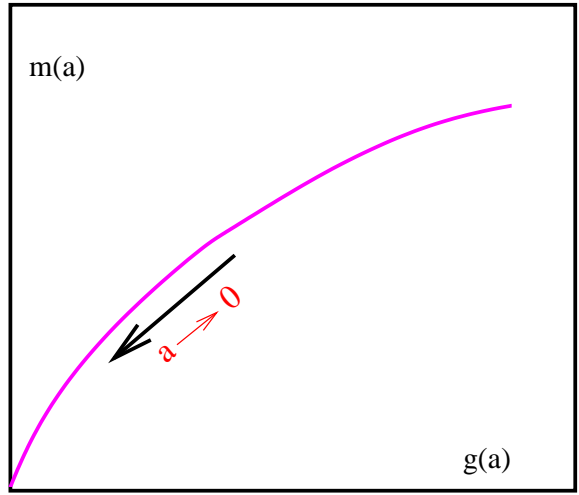}
\caption{The continuum limit involves following a renormalization
group trajectory in the space of bare parameters.}
\label{rgflow}
\end{figure}

At a basic level these equations arise from holding a few physical
quantities fixed along the ``renormalization group trajectory,'' as
sketched in Fig.~\ref{rgflow}.  For this discussion it is convenient
to use the lightest baryon and the lightest meson masses, $m_p$,
$m_\pi$, as physical quantities to hold fixed.  With multiple quark
flavors one would hold several meson masses fixed.

The parameters $\Lambda$ and $M$ represent ``integration constants''
of the renormalization group equations.  $\Lambda$ is conventionally
interpreted as the ``QCD scale,'' while $M$ defines a the
``renormalized quark mass.''  Their values follow from limits along
the renormalization group trajectory
\begin{equation}
\Lambda=\lim_{a\rightarrow 0} \ { e^{-1/2\beta_0 g^2}
g^{-\beta_1/\beta_0^2}\over a}
\end{equation}
\bigskip
\begin{equation}
M=\lim_{a\rightarrow 0}\ m g^{-\gamma_0/\beta_0}
\end{equation}
\bigskip
The precise numerical values of $\Lambda$ and $M$ depend on the
renormalization scheme.

The physical masses map directly onto the integration constants,
$\Lambda=\Lambda(m_p,m_\pi)$ and $M=M(m_p,m_\pi)$.  Inverting, gives
the physical masses as functions of $\Lambda$ and $M$:
$\longrightarrow$ $m_i=m_i(\Lambda,M)$.  Simple dimensional analysis
implies this relationship must take the form $m_i=\Lambda
f_i(M/\Lambda)$, with $f(x)$ some {\it a priori} unknown function.

In the case of multiple degenerate fermions more is known about the
behavior of $f$. In particular, Goldstone bosons should appear as the
quark mass goes to zero: $m_\pi^2 \sim m_q$.  This implies the
existence of a square root singularity {$f_\pi(x)\sim x^{1/2}$}.  The
location of this singularity defines what is meant by zero mass
quarks, thus removing any additive ambiguity in defining $M$.

The single massless flavor case, however, is somewhat special.  Then
the meson mass $m_\pi=\Lambda f_\pi(M/\Lambda)$ does not vanish at
$M=0$.  The anomaly precludes chiral symmetry and Goldstone bosons.
While the meson mass can be forced to vanish, the earlier discussion
shows that this requires a special tuning to a negative quark mass.
For the function $f_\pi(x)$ one expects a smooth and non-vanishing
behavior at $x=0$.

\begin{figure}
  \includegraphics[height=.3\textheight]{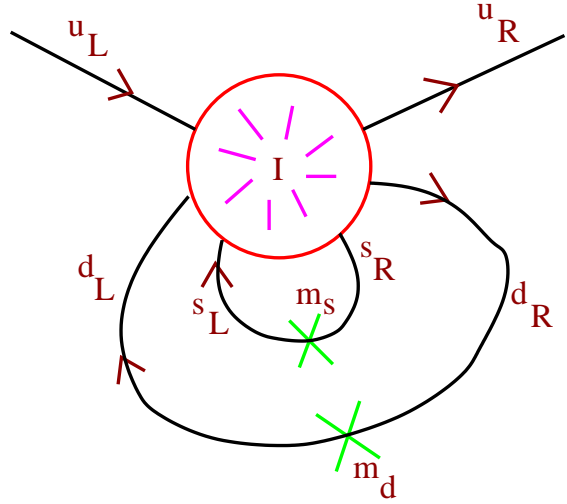}
\caption{Non-perterbative effects can flip all quark spins through
the anomaly.  Tying the heavier quark lines together with mass terms
generates a shift in the lightest quark mass.}
\label{thooft}
\end{figure}

The shift of the singularity in $x$ away from zero is due to
non-perturbative effects.  Indeed, non-perturbative contributions to
the mass flow are expected which are not proportional to quark mass.
As shown some time ago by t'Hooft \cite{'tHooft:fv}, there are
nonperturbative classical effects called ``instantons'' that flip all
quark spins simultaneously.  Tying the heavier quarks together with
their mass terms as sketched in Fig.~{\ref{thooft}} gives rise to a
contribution to the up quark mass flow
\begin{equation}
\Delta m_u \sim {m_d m_s\over\Lambda_{\rm qcd}},\ \ \Lambda_{\rm qcd}
\end{equation}
These effects indicate that $m_u=0$ is not renormalization group
invariant.  If the up quark mass vanishes as some point on the
trajectory, it will not for other points.

\subsection{Matching between schemes}

This non-invariance of the quark mass under the renormalization group
raises the specter of scheme dependence.  On changing renormalization
schemes one should preserve the lowest order perturbative limit as
$g\rightarrow 0$ at fixed scale $a$
\begin{equation}\matrix{
&\tilde g= g+O(g^3)+\hbox{ non-perturbative}\cr
&\cr
&\tilde m= m(1+O(g^2))+\hbox{ non-perturbative}\cr
}
\end{equation}
Here ``non-perturbative'' terms should vanish faster than any power of
$g$.  As mentioned above, the integration constants $\Lambda$ and $M$
in general depend on the chosen scheme.

It is important to recognize that this perturbative matching at fixed
$a$ is not the continuum limit.  Indeed, taking $g\rightarrow 0$ at
fixed $a$ gives perturbation theory on free quarks, while taking
$a\rightarrow 0$ at fixed $g$ gives the divergences of field theory.
For the confining physics of the real world, one must go between these
limits and take $a$ and $g$ together to zero along the renormalization
group trajectory.

To dramatically illustrate the issue, consider a particularly cooked
up new scheme
\begin{equation}\matrix{
\tilde a = a\cr
\tilde g = g\cr
\tilde m=m-M g^{\gamma_0/\beta_0}\times
{ e^{-1/2\beta_0 g^2} g^{-\beta_1/\beta_0^2}\over \Lambda a}\cr
}
\end{equation}
Here I have crafted the last factor in the mass expression to approach
unity.  This non-perturbative redefinition of the bare parameters
makes
\begin{equation}
\tilde M\equiv
\lim_{a\rightarrow 0} \tilde m \tilde g^{-\gamma_0/\beta_0} = 
M-M=0
\end{equation}
While this may be somewhat artificial, it shows that some scheme
always exists where the renormalized quark mass vanishes.  Of course
doing this for the top quark will insert ridiculously large
non-perturbative effects, but it is possible in principle.  Because of
this ambiguity, $M=0$ is scheme dependent and thus is not a physical
concept.  Of course with degenerate quarks one can precisely define
masslessness by the location of the square root singularity in $f(x)$
as defined above.

\subsection{On the lattice}

As in the general case, on the lattice the renormalization flows
depend on details of the lattice action.  Various gauge actions as
well as the fermion formulation need to be considered.  Recent
discussions have concentrated on overlap/Ginsparg-Wilson fermion
operators \cite{Neuberger:1997fp,Ginsparg:1981bj}, which bring a
remnant of chiral symmetry to the lattice.  However even these
operators are not unique.  The overlap operator relies on a
projection, but it depends on the particular Dirac operator being
projected.  When starting with a Wilson Dirac operator, the input
negative mass is adjustable over a finite range.

The one flavor theory dynamically generates a gap which will appear in
the spectrum of the final Dirac operator.  The overlap projection does
not protect the size of this gap.  With the gap present, the
Ginsparg-Wilson condition is not sufficient to guarantee the
preservation of $M=0$ between schemes.

With a Ginsparg-Wilson action the concept of a massless quark is
synonymous with zero topological susceptibility.  If a single massless
quark is an ill defined concept, this raises the question of whether
the topological susceptibility is uniquely defined for $N_f<2$.  This
question has also been asked in the context of making the
Witten-Veneziano formula \cite{Witten:1979vv,Veneziano:1979ec} for the
$\eta^\prime$ mass precise \cite{Seiler:1987ig,Seiler:2001je}.  From a
perturbative point of view infinities are not a problem
\cite{Giusti:2004qd,Luscher:2004fu}.  However, to give a unique
winding number to a gauge configuration requires a degree of
smoothness for the gauge fields.  One popular condition that ensures a
well defined winding number forbids plaquettes further than a finite
distance $\delta$ from the identity \cite{Luscher:1981zq}.  However
this condition is rather strong and leaves unresolved issues.  In
particular this ``admissibility condition'' has recently been shown to
be in conflict with reflection positivity \cite{Creutz:2004ir}.

\section{CONCLUSIONS}

Effective chiral Lagrangians show that the strong interactions will
spontaneously violate CP for large regions of parameter space.  This
phenomenon requires negative quark masses, a concept made physical by
the anomaly.  Based on this qualitative picture, I have argued that
$m_u=0$ is not a meaningful concept.  As such, it cannot be regarded
as a possible solution to the strong CP problem.  These effects are
entirely non-perturbative.  As a corollary, the topological
susceptibility is not uniquely defined for $N_f<2$.  Finally, I note
that available simulation algorithms cannot explore this fascinating
physics because it involves regions where the fermion determinant is
not positive, i.e. Monte Carlo methods have a sign problem.

%%%%%%%%%%%%%%%%%%%%%%%%%%%%%%%%%%%%%%%%%%%%%%%%
%% BACKMATTER
%%%%%%%%%%%%%%%%%%%%%%%%%%%%%%%%%%%%%%%%%%%%%%%%

\begin{theacknowledgments}
This manuscript has been authored under contract number
DE-AC02-98CH10886 with the U.S.~Department of Energy.  Accordingly,
the U.S. Government retains a non-exclusive, royalty-free license to
publish or reproduce the published form of this contribution, or allow
others to do so, for U.S.~Government purposes.
\end{theacknowledgments}

\end{document}